%% file: main.tex
\def\arxiv{true}
\documentclass[sigconf]{acmart}
\input{dlab_macros}
 
\usepackage{graphicx} 
\usepackage{balance}
\usepackage{amsfonts,amsmath,amsthm}
\usepackage{bm}
\usepackage{booktabs} 
\usepackage{threeparttable}
\usepackage[htt]{hyphenat}
\usepackage{enumitem}
\usepackage{xspace}
\usepackage{multirow}
\usepackage{makecell}
\usepackage{hhline}
\usepackage{hyperref}
\usepackage{url}
\usepackage{colortbl}
\usepackage{subfig}
\usepackage[normalem]{ulem} 
\usepackage{algorithm}
\usepackage{appendix}   
\usepackage[noend]{algorithmic}
\usepackage{arydshln}

\setcitestyle{numbers,super&compress}

\newcommand{\ignore}[1]{}

\newcommand{\tabcaption}[1]{\vspace*{-3mm}\caption{#1}\vspace*{-4mm}}
\newcommand{\figcaption}[1]{\vspace*{-3mm}\caption{#1}\vspace*{-5mm}}
\newcommand{\moveup}{\vspace*{-2mm}}
\newcommand{\moveups}{\vspace*{-1mm}}

\newcommand{\martin}[1]{\textcolor{black}{#1}}

\newcommand{\logs}{\textsc{Logs}\xspace}
\newcommand{\cspriv}{\textsc{Clickstream-Priv}\xspace}
\newcommand{\cspub}{\textsc{Clickstream-Pub}\xspace}

\newcommand{\graph}{\textsc{Graph}\xspace}

\newcommand{\fscore}{$F_1$-score\xspace}

\newcommand{\sklearn}{\textsc{Scikit-learn}\xspace}

\newcommand{\en}{\textsc{English}\xspace}
\newcommand{\ja}{\textsc{Japanese}\xspace}
\newcommand{\de}{\textsc{German}\xspace}
\newcommand{\ru}{\textsc{Russian}\xspace}
\newcommand{\fr}{\textsc{French}\xspace}
\newcommand{\itl}{\textsc{Italian}\xspace}
\newcommand{\pl}{\textsc{Polish}\xspace}
\newcommand{\fa}{\textsc{Persian}\xspace}

\newcommand{\enshort}{\textsc{EN}\xspace}
\newcommand{\jashort}{\textsc{JA}\xspace}
\newcommand{\deshort}{\textsc{DE}\xspace}
\newcommand{\rushort}{\textsc{RU}\xspace}
\newcommand{\frshort}{\textsc{FR}\xspace}
\newcommand{\itlshort}{\textsc{IT}\xspace}
\newcommand{\plshort}{\textsc{PL}\xspace}
\newcommand{\fashort}{\textsc{FA}\xspace}

\ifx\arxiv\undefined
\else
\fi

\copyrightyear{2022}
\acmYear{2022}
\setcopyright{acmlicensed}

\ifx\arxiv\undefined
  \acmConference[WSDM '22] {Proceedings of the Fifteenth ACM International Conference on Web Search and Data Mining}{February 21--25, 2022}{Tempe, AZ, USA.}
  \acmBooktitle{Proceedings of the Fifteenth ACM International Conference on Web Search and Data Mining (WSDM '22), February 21--25, 2022, Tempe, AZ, USA}
  \acmPrice{15.00}
  \acmISBN{978-1-4503-9132-0/22/02}
  \acmDOI{10.1145/3488560.3498496}
\else
  \acmConference[arXiv]{arXiv e-print}
  \acmBooktitle{}
  \acmPrice{}
  \acmISBN{}
  \acmDOI{10.1145/3488560.3498496}
\fi

\begin{document}

\title{Wikipedia Reader Navigation: When Synthetic Data Is Enough}

\author{Akhil Arora}
\affiliation{%
  \institution{EPFL}
  \country{}
}
\email{akhil.arora@epfl.ch}

\author{Martin Gerlach}
\affiliation{%
  \institution{Wikimedia Foundation}
  \country{}
}
\email{mgerlach@wikimedia.org}

\author{Tiziano Piccardi}
\affiliation{%
  \institution{EPFL}
  \country{}
}
\email{tiziano.piccardi@epfl.ch}

\author{Alberto Garc\'{i}a-Dur\'{a}n}
\authornote{Research done while at EPFL.}
\affiliation{%
  \institution{Atinary Technologies}
  \country{}
}
\email{agaduran@gmail.com}

\author{Ro\-bert West}
\authornote{Robert West is a Wikimedia Foundation Research Fellow.}
\affiliation{%
  \institution{EPFL}
  \country{}
}
\email{robert.west@epfl.ch}

\renewcommand{\shortauthors}{Arora \etal}

\begin{abstract}
Every day millions of people read Wikipedia.
When navigating the vast space of available topics using hyperlinks, readers describe trajectories on the article network.
Understanding these navigation patterns is crucial to better serve readers' needs and address structural biases and knowledge gaps. 
However, systematic studies of navigation on Wikipedia are hindered by a lack of publicly available data due to the commitment to protect readers' privacy by not storing or sharing potentially sensitive data. 
In this paper, we ask: How well can Wikipedia readers' navigation be approximated by using publicly available resources, most notably the Wikipedia clickstream data? 
We systematically quantify the differences between real navigation sequences and synthetic sequences generated from the clickstream data, in 6 analyses across 8 Wikipedia language versions. 
Overall, we find that the differences between real and synthetic sequences are statistically significant, but with small effect sizes, often well below $10\%$. 
This constitutes quantitative evidence for the utility of the Wikipedia clickstream data as a public resource: clickstream data can closely capture reader navigation on Wikipedia \martin{and provides a sufficient approximation for most practical downstream applications relying on reader data}. 
More broadly, this study provides an example for how clickstream-like data can generally enable research on user navigation on online platforms while protecting users' privacy.
\end{abstract}

\maketitle

\input{intro}
\input{rwork}
\input{problem}
\input{methods}
\input{analysis}
\input{downstream_tasks}
\input{discussion}

\section{Conclusions}
In this study, we addressed the question of how much of the complexity of reader navigation in Wikipedia can be captured by the publicly available clickstream data.
We systematically compared ensembles of real and synthetically generated navigation sequences across 8 different languages.
We found that differences are statistically significant but absolute effect sizes are small establishing quantitative confidence in the utility of clickstream data to study navigation more generally.
Our results empower more researchers to study navigation in Wikipedia and further strengthen the privacy of readers by limiting the need to share further potentially sensitive data in most practical use-cases. 

\begin{acks}
We thank Leila Zia for insightful discussions. This project was partly funded by the Swiss National Science Foundation (grant 200021\_185043), the European Union (TAILOR, grant 952215), and the Microsoft Swiss Joint Research Center. We also gratefully acknowledge generous gifts from Facebook and Google supporting West’s lab.
\end{acks}

{
\newpage
\bibliographystyle{ACM-Reference-Format}
\balance
\bibliography{nav}
}

\setcounter{figure}{0}
\setcounter{table}{0}
\setcounter{section}{0}
\setcounter{subsection}{0}
\makeatletter
\renewcommand{\thefigure}{S\arabic{figure}}
\renewcommand{\thetable}{S\arabic{table}}

\begin{figure*}[!htb]
  \moveups
    \includegraphics[width=0.99\linewidth]{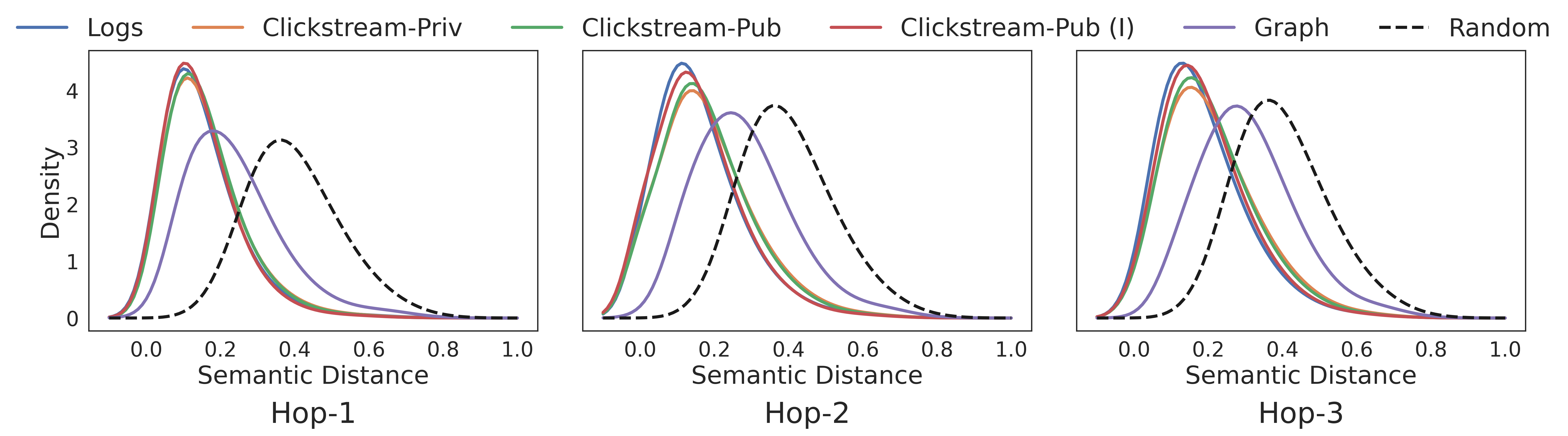}
    \caption{Diffusion in semantic space: Distribution of cosine distance between the first article of a sequence and the article after $k$-steps ($k=1,2,3$) for \en Wikipedia.}
    \label{fig:semantic_diffusion_app}
\end{figure*}

\appendixpage
\input{appendix}
\end{document}

%% file: dlab_macros.tex
\usepackage{xspace}
\usepackage{amsmath}
\usepackage{amsfonts}
\usepackage{url}
\usepackage{marvosym}
\usepackage{ifthen}
\theoremstyle{plain}

\newcommand{\chatoDisplayMode}[1]{#1}

\definecolor{MyRed}{rgb}{0.6,0.0,0.0} 
\definecolor{MyBlack}{rgb}{0.1,0.1,0.1} 
\newcommand{\inred}[1]{{\color{MyRed}\sf\textbf{\textsc{#1}}}}
\newcommand{\frameit}[2]{
  \begin{center}
  {\color{MyRed}
  \framebox[.9\columnwidth][l]{
    \begin{minipage}{.85\columnwidth}
    \inred{#1}: {\sf\color{MyBlack}#2}
    \end{minipage}
  }\\
  }
  \end{center}
}

\newcommand{\note}[2][]{\chatoDisplayMode{\def\@tmpsig{#1}\frameit{{\Pointinghand} Note}{#2\ifx \@tmpsig \@empty \else \mbox{ --\em #1}\fi}}}
\newcommand{\todo}[2][]{\chatoDisplayMode{\def\@tmpsig{#1}\frameit{{\Writinghand} To-do}{#2\ifx \@tmpsig \@empty \else \mbox{ --\em #1}\fi}}}

\newcommand{\abbrevStyle}[1]{#1}

\newcommand{\ie}{\abbrevStyle{i.e.}\xspace}
\newcommand{\eg}{\abbrevStyle{e.g.}\xspace}
\newcommand{\cf}{\abbrevStyle{cf.}\xspace}
\newcommand{\etal}{\abbrevStyle{et al.}\xspace}

\newcommand{\etc}{\abbrevStyle{etc.}\xspace}

\newcommand{\Figref}[1]{Fig.~\ref{#1}}

\newcommand{\xhdr}[1]{\vspace{1.7mm}\noindent{{\bf #1.}}}

\newcommand{\textcite}[1]{\citeauthor{#1} \shortcite{#1}}

\newcommand{\hide}[1]{}

\makeatletter
\newcommand{\iffont}[2]{\ifthenelse{\equal{\f@family}{#1}}{#2}{}}
\makeatother

\iffont{ptm}{
  \usepackage{mathptmx}

  \DeclareSymbolFont{greek}{OML}{cmm}{m}{n}
  \DeclareMathSymbol{\alpha}{\mathalpha}{greek}{"0B}
  \DeclareMathSymbol{\beta}{\mathalpha}{greek}{"0C}
  \DeclareMathSymbol{\gamma}{\mathalpha}{greek}{"0D}
  \DeclareMathSymbol{\delta}{\mathalpha}{greek}{"0E}
  \DeclareMathSymbol{\epsilon}{\mathalpha}{greek}{"0F}
  \DeclareMathSymbol{\zeta}{\mathalpha}{greek}{"10}
  \DeclareMathSymbol{\eta}{\mathalpha}{greek}{"11}
  \DeclareMathSymbol{\theta}{\mathalpha}{greek}{"12}
  \DeclareMathSymbol{\iota}{\mathalpha}{greek}{"13}
  \DeclareMathSymbol{\kappa}{\mathalpha}{greek}{"14}
  \DeclareMathSymbol{\lambda}{\mathalpha}{greek}{"15}
  \DeclareMathSymbol{\mu}{\mathalpha}{greek}{"16}
  \DeclareMathSymbol{\nu}{\mathalpha}{greek}{"17}
  \DeclareMathSymbol{\xi}{\mathalpha}{greek}{"18}
  \DeclareMathSymbol{\pi}{\mathalpha}{greek}{"19}
  \DeclareMathSymbol{\rho}{\mathalpha}{greek}{"1A}
  \DeclareMathSymbol{\sigma}{\mathalpha}{greek}{"1B}
  \DeclareMathSymbol{\tau}{\mathalpha}{greek}{"1C}
  \DeclareMathSymbol{\upsilon}{\mathalpha}{greek}{"1D}
  \DeclareMathSymbol{\phi}{\mathalpha}{greek}{"1E}
  \DeclareMathSymbol{\chi}{\mathalpha}{greek}{"1F}
  \DeclareMathSymbol{\psi}{\mathalpha}{greek}{"20}
  \DeclareMathSymbol{\omega}{\mathalpha}{greek}{"21}
  \DeclareMathSymbol{\varepsilon}{\mathalpha}{greek}{"22}
  \DeclareMathSymbol{\vartheta}{\mathalpha}{greek}{"23}
  \DeclareMathSymbol{\varpi}{\mathalpha}{greek}{"24}
  \DeclareMathSymbol{\varrho}{\mathalpha}{greek}{"25}
  \DeclareMathSymbol{\varsigma}{\mathalpha}{greek}{"26}
  \DeclareMathSymbol{\varphi}{\mathalpha}{greek}{"27}
  \DeclareSymbolFont{otone}{OT1}{cmr}{m}{n}
  \DeclareMathSymbol{\Gamma}{\mathalpha}{otone}{0}
  \DeclareMathSymbol{\Delta}{\mathalpha}{otone}{1}
  \DeclareMathSymbol{\Theta}{\mathalpha}{otone}{2}
  \DeclareMathSymbol{\Lambda}{\mathalpha}{otone}{3}
  \DeclareMathSymbol{\Xi}{\mathalpha}{otone}{4}
  \DeclareMathSymbol{\Pi}{\mathalpha}{otone}{5}
  \DeclareMathSymbol{\Sigma}{\mathalpha}{otone}{6}
  \DeclareMathSymbol{\Upsilon}{\mathalpha}{otone}{7}
  \DeclareMathSymbol{\Phi}{\mathalpha}{otone}{8}
  \DeclareMathSymbol{\Psi}{\mathalpha}{otone}{9}
  \DeclareMathSymbol{\Omega}{\mathalpha}{otone}{10}
  \DeclareSymbolFont{syms}{OML}{cmm}{m}{it}
  \DeclareMathSymbol{\partial}{\mathord}{syms}{"40}
  \DeclareMathAlphabet{\mathbold}{OML}{cmm}{b}{it}
  \DeclareSymbolFont{largesymbols}{OMX}{cmex}{m}{n}

}

%% file: intro.tex
\section{Introduction}
\label{sec:intro}

Wikipedia has become one of the largest platforms for open knowledge in the world with more than 56M articles in 300+ languages and approximately 250K new articles added each month. 
However, the knowledge is not just the sum of the individual articles. 
In fact, articles in Wikipedia are connected via an extensive network of hyperlinks; \eg, English Wikipedia's roughly 6M articles are connected by more than 400M hyperlinks \cite{Mitrevski2020wikihist}. These hyperlinks allow readers to navigate from one page to another, find related information, or discover new topics.

But how do readers use this network to navigate the information contained in Wikipedia? Insights into patterns of navigation are extremely important for identifying content that is missing or hard-to-find in order to address knowledge gaps~\cite{Zia2019knowledge}. 
For example, the gender gap in Wikipedia, which is commonly associated with an under-representation of biographies on women~\cite{Redi2020taxonomy}, is also reflected in structural properties such as lower centrality and visibility in the network 
~\cite{Graells-Garrido2015first, Langrock2020gender}.
This in turn has been shown to lead to biases in the navigation of readers~\cite{Menghini2020wikipedia}, adding to observations on gender disparities in Wikipedia readership~\cite{Johnson2021global}.
A better understanding of navigation in this context could also help empower readers to find content more easily, or to organize lists of articles when learning about a specific topic in the form of a curriculum~\cite{Sayyadiharikandeh2019finding}.

The phenomenon of navigating through different topics when browsing Wikipedia has been colloquially described as a ``rabbit hole'' \cite{wiki_rabit_hole1, wiki_rabit_hole2}. However, in quantitative terms, our understanding of the learning pathways that readers are actually following is still extremely limited. 
There have been extensive studies of navigation on Wikipedia in game-like settings (\eg Wikigame or Wikispeedia), where the aim is to find paths between two given articles~\cite{HumanWayfinding}. 
However, \martin{results from targeted navigation cannot be expected to apply in the same way to “navigation in the wild”} \cite{piccardi2021largescale}.

One of the main limitations to address these questions is the lack of publicly available data. 
The raw webrequest logs~\cite{webrequest_meta}, 
which capture all requests to Wikimedia’s servers (including IP addresses), are considered extremely sensitive.
The Wikimedia Foundation's commitment to protect readers' privacy~\cite{wmf_privacy_policy} requires this data to remain private and, in fact, to be completely purged after 90 days.
In an effort to accommodate research interests, anonymized and/or aggregated derived datasets are kept for longer time spans and shared publicly.
For example, the pageview dumps~\cite{wikipedia_full_pageview_dumps} provide timeseries of the number of pageviews per article per hour. 
These provide some proxy for article popularity (see, \eg, the pageviews tool~\cite{wikipedia_pageviews_tool}) as well as collective attentions more generally \cite{Ribeiro2020sudden}. 

In the same vein, the clickstream dataset~\cite{clickstream_dumps} has been released as a public resource capturing some aspects of navigation on Wikipedia. It contains the counts of source--target pairs of articles (\ie, how often a link from one page to another page was clicked by readers) in a given month. It has not only been used to study the popularity of links \cite{LinkSuccessfulWikipedia}, but also for generating synthetic reading pathways to infer properties of navigation patterns \cite{SearchStrategies}.

Here, for the first time, we address the question of how much of the complexity of reader navigation on Wikipedia can be captured by the publicly available clickstream data.
More precisely, we systematically compare ensembles of real and synthetically generated navigation sequences via two studies empirically characterizing the sequences and four downstream tasks across eight different languages.
We find that differences are statistically significant but absolute effect sizes are small, establishing quantitative confidence in the utility of clickstream data to study navigation more generally.

%% file: rwork.tex
\section{Related Work}
\label{sec:rwork}
In this section, we review works studying readers on Wikipedia and their navigation, as well as navigation on the web more generally.

\xhdr{Wikipedia readers' pageviews}
Wikipedia's publicly available pageviews-data~\cite{wikipedia_pageview_dumps} (i.e. how often individual pages are viewed) serves as a common proxy to measure collective attention to, e.g., news or popular topics~\cite{Garcia-Gavilanes2017memory, Ribeiro2020sudden}.
Other use cases involve the detection of misalignment between supply and demand to content by comparing pageviews to quality ratings~\cite{Warncke-Wang2015misalignment}.
Furthermore, individual studies have measured the dwell-time, that is how much time readers actually spend on a given page, finding substantial differences across geographic regions~\cite{DwellingTime}. 
More generally, \citeauthor{Shaw2018pipeline}~\cite{Shaw2018pipeline} propose a model for online participation in the form of a pipeline, where being a reader is one of many steps for a user to become an active contributor.
Thus, readers have been considered a core-dimension in recent efforts to systematically approach knowledge gaps in Wikipedia~\cite{Redi2020taxonomy}.

\xhdr{Navigation on Wikipedia}
Only few studies have characterized the navigation of readers on Wikipedia since data from webrequest-logs is not public.   
Halfaker \etal~\cite{InterActivityTime} analyzed the distribution of inter-activity times (i.e. the time between two pageviews by the same user) to determine a threshold when constructing reading sessions.
Lehmann \etal~\cite{ReaderPreferences} identify four different types of reading patterns by performing a k-means clustering of reading sessions obtained (with consent) from anonymized activity log data via the Yahoo-toolbar.
Lydon-Staley \etal~\cite{Lydon-Staley2021} found $2$ different types of curiosity-seeking behavior when looking at the knowledge networks created from the browsing sessions of $\approx 150$ volunteers.
Two recent studies investigated why readers visit Wikipedia~\cite{singer_why_2017, lemmerich_why_2019} and correlated responses to surveys on motivation, information need, and prior knowledge with features from reading sessions such as session length.
Paranjape \etal~\cite{ParanjapeImproving} showed that navigation traces provide a strong signal to predict new useful links among Wikipedia articles.
Furthermore, reading sessions can be used to construct article embeddings---where similarity captures articles that are read in close succession~\cite{wiki_nav_vectors}---that have been used for building alternative visual representations of the semantic space of Wikipedia~\cite{Sen2019towards}.

\xhdr{Applications of clickstream data}
Due to lack of availability of data on Wikipedia's webrequest-logs, several studies have used the clickstream dataset as a proxy to characterize navigation. While the latter contains only aggregate counts of the number of times a given link from a source- to a target-page was clicked, these studies already provide important insights into readers' behavior.
While generally only a small fraction of existing links is used, it was found that popularity of links depends strongly on different properties of the article (e.g. in-/out-degree, centrality) or the placement of links within the article (e.g. whether they occur in the lead section)~\cite{LinkSuccessfulWikipedia,StructureArticlesNavigation}. 
Comparing the incoming and outgoing clicks of a page allows for a classification of pages based on usage, e.g. into sources/sink or bottlenecks~\cite{Gildersleve2018inspiration,dimitrov2019different}. 
Other approaches used the clickstream data to assess whether some articles should be read before others when learning about a specific topic~\cite{Sayyadiharikandeh2019finding}, to detect structural biases in content ~\cite{Menghini2020wikipedia}, extract semantic relationships between articles~\cite{Dallmann2016extracting}, as a ground-truth for other tasks~\cite{Schwarzer2016evaluating,Consonni2020cyclerank}, or to study how the structure of the page influences the links clicked by the readers \cite{VisualPositions}.
Finally, Rodi \etal~\cite{SearchStrategies} characterized search strategies in synthetic data generated from clickstream, however, not without  making strong assumptions about the underlying process of navigation.

\xhdr{Targeted navigation}
The most detailed studies on navigation on Wikipedia come from experiments such as Wikigame~\cite{wikigame} or Wikispeedia~\cite{wikispeedia_game,Wikispeedia}, where participants are asked to find a path between two given articles using the hyperlinks. 
Several studies developed models to assess the order of the Markov process~\cite{Singer2014detecting,Petrovic2020learning}, or assess other hypothesis about the navigation (such as decentralized search)~\cite{Helic2013models,Singer2015hyptrails}. 
Empirical approaches aimed at characterizing different properties of the navigation traces, such as the evolution of step sizes~\cite{HumanWayfinding} or why users stop~\cite{LastClick}. 
\martin{However, conclusions from studies on targeted navigation do not apply in the same way to ``navigation in the wild'' \cite{piccardi2021largescale}.}

\xhdr{Navigation in general}
Understanding navigation on the web and online platforms has been a long-standing problem~\cite{Catledge1995characterizing}. 
Some of the common approaches include, e.g., identification of specific patterns~\cite{RevisitationWWWNavigation,UserConsumption}, development of generative models~\cite{UsersReallyMarkovian,RandomSurfers}, or the characterization of the  overall predictability~\cite{Kulshrestha2020web}.

%% file: problem.tex
\section{Problem and Notation}
\label{sec:problem}
In this section, we formalize the problem of assessing the utility of the public Wikipedia clickstream data~\cite{clickstream_dumps} to characterize the navigation of readers in Wikipedia.

We start from the set of real navigation sequences containing the sequences of pageviews by the same user (indexed by $i$), 
\begin{equation}
\mathcal{S} = \left\{ (X_1^{(i)},X_2^{(i)},\ldots,X_{n_i}^{(i)})\right\}_{i=1,\ldots,N},  
\end{equation}
with $n_i$ being the length (or the number of pageviews) of the sequence $i$.
Next, we generate a set of synthetic navigation sequences based on the clickstream data with similar properties, 
\begin{equation}
\hat{\mathcal{S}} = \left\{ (X_1^{(i)},\hat{X}_2^{(i)},\ldots,\hat{X}_{n_i}^{(i)})\right\}_{i=1,\ldots,N}, 
\end{equation}
where $\hat{X}_j^{(i)}$ with $j=2,\ldots,n_i$ are pageviews generated via a random walk on the network of Wikipedia articles.

Given the sets $\mathcal{S}$ and $\hat{\mathcal{S}}$ of navigation sequences, we conduct multiple analyses and experiments: i) summary statistics derived from empirical characterization of the sequences (\cf Sec.~\ref{sec:approach}), and ii) performance-metrics in downstream tasks (\cf Sec.~\ref{sec:tasks}), and leverage the observables to quantify the difference between $\mathcal{S}$ and $\hat{\mathcal{S}}$.

%% file: methods.tex
\section{Data and Resources}
\label{sec:data}
In this section, we describe in detail the datasets used for the analysis of reader navigation in Wikipedia. 
We consider 8 different language versions of Wikipedia for which clickstream data is available, namely---\en (\enshort), \ja (\jashort), \de (\deshort), \ru (\rushort), \fr (\frshort), \itl (\itlshort), \pl (\plshort), and \fa (\fashort). Additionally, this choice covers languages from different families.
The dataset statistics are presented in Table~\ref{tab:data_stats_app}. 
\martin{All the publicly accessible resources required to reproduce the experiments in this paper are available at~\url{https://github.com/epfl-dlab/wikinav-approx}}.

Next, we describe the procedure for constructing the five different types (\cf Table~\ref{tab:data_overview}) of navigation sequences analyzed in this work, as well as all auxiliary resources.

\subsection{Real navigation sequences}
\label{sec:real_nav}
We construct navigation sequences for readers of Wikipedia articles using Wikipedia's Webrequest server logs from March, 2021. We only consider internal navigation, \ie, navigation to/from websites external to Wikipedia are ignored. 

\xhdr{Webrequest logs~\cite{webrequest_meta}} The webrequest logs contain an entry for each HTTP request to Wikimedia servers, specifying information, including, but not limited to, timestamp, requested URL, referrer URL, client IP address, and user agent information.

\xhdr{Constructing navigation sequences} 
Real navigation sequences are constructed following the approach proposed by \citeauthor{ParanjapeImproving} \cite{ParanjapeImproving}. 
As a first step, we construct navigation sessions for each Wikipedia reader. Note that a reader may visit multiple pages from a given page by opening different browser tabs, and thus, navigation sessions are innately trees and not chains. 
We construct navigation trees by stitching together pageviews using the referrer information. 
From this, we obtain navigation sequences by sampling, uniformly at random, one root-to-leaf path from each navigation tree with at least 2 nodes. 
Note that we do not consider all root-to-leaf paths in a tree, as that would lead to navigation sessions with more complex trees to be over-represented when compared to those with a simple tree. 
We observed that most sequences are short, however, long sequences do exist. For instance, $75\%$ ($600M$) of the sequences in the \en Wikipedia are of length $2$, but only $0.5\%$ ($4M$) are of length $10$ or more. 

\subsection{Synthetic navigation sequences}
\label{sec:synthetic_gen}
We construct synthetic navigation sequences using the Wikipedia clickstream and the hyperlink graph from March, 2021.

\xhdr{Wikipedia hyperlink graph}
We construct the `directed' hyperlink graph from the links contained in the main text of Wikipedia articles available via the XML dumps~\cite{wikipedia_xmldumps} released by Wikipedia on a monthly basis. We perform link extraction based on the approach described in \cite{link_extraction_wikitext}. To align the article versions with the webrequest logs and the clickstream data from March 2021, we use the dumps dated `2021-04-01'.

\xhdr{Clickstream dataset}
The Wikipedia clickstream~\cite{Wulczyn2015clickstream} contains the counts of links (i.e. pairs of source-target pages) clicked by Wikipedia readers. The data is generated from the webrequest logs, and is published on a monthly basis as dump files~\cite{clickstream_dumps} for $11$ different language versions of Wikipedia. There are several pre-processing steps and filters; most importantly, the removal of links with $10$ or fewer observations for protecting the privacy of readers.

\begin{table}[t]
\centering
\normalsize
\vspace{2mm}
\tabcaption{\label{tab:data_overview} An overview of the type of navigation sequences analyzed in this work.}
\resizebox{1\columnwidth}{!}{
    \begin{tabular}{lll}
    \toprule
     \textbf{Dataset} & \textbf{Type} & \textbf{Main Characteristics}  \\ \midrule  
     \logs & Real & Human navigation on Wikipedia. \\ \hdashline
     \cspriv & Synthetic & Markov-1, biased random walks using private Clickstream. \\
     \cspub & Synthetic & Markov-1, biased random walks using public Clickstream. \\
     \multirow{2}{*}{\cspub (I)} & \multirow{2}{*}{Synthetic} & \martin{Markov-1, biased random walks using public Clickstream,} \\
      & & \martin{with a different intrinsic stopping criterion \cite{SearchStrategies}.} \\
     \graph & Synthetic & Markov-1, unbiased random walks on Wikipedia hyperlink graph. \\
        \bottomrule
    \end{tabular}}
\moveup
\moveup
\end{table}

\xhdr{Generating Markovian navigation sequences} 
We generate four different types of synthetic navigation sequences closely mimicking the overall statistics of the real sequences, by ensuring: i) for each generated sequence the same starting article and length\footnote{If the random walker reaches an isolated node prior to reaching the desired sequence length, we back-track and restart the walk from the parent. This step is performed until the desired length is reached.} as its corresponding real sequence, and ii) the number of generated sequences to be the same as the total number of real sequences. With these constraints in place, we generate sequences by performing random walks with the following three different transition probabilities between articles in the Wikipedia hyperlink graph:
\begin{itemize}[leftmargin=*, noitemsep]
    \item \textbf{\cspriv:} weights proportional to the number of clicks between two articles obtained from the webrequest logs, which is similar to \cspub, except that it also includes pairs of articles with $10$ or fewer observations.
    \item \textbf{\cspub:} weights proportional to the number of clicks between two articles obtained from the publicly available clickstream data.
    \item \textbf{\graph:} weights are uniform over all outgoing links from the Wikipedia hyperlink graph.
\end{itemize}
\martin{Note that \cspub (I) uses the same transition probabilities as \cspub. The only difference is that the former not only decides the next step but also takes the decision to stop at a given node based on the pairwise transition probabilities~\cite{SearchStrategies} (``intrinsic stopping''). Thus, unlike other strategies that use extrinsic stopping, the length constraint is not enforced for navigation sequences obtained using \cspub (I).} Additionally, note that the state space for the random walker is all the articles in the Wikipedia hyperlink graph and not just the ones observed in the webrequest logs or clickstream. While the unobserved articles would never be visited via the weighted random walker, the unbiased random walker would visit them.

\subsection{Auxiliary resources}
\label{sec:data_auxiliary}
For the empirical analysis and downstream tasks we use multiple additional resources, which are described as follows.

\begin{figure*}[!htb]
  \moveups
  \includegraphics[width=0.3\linewidth]{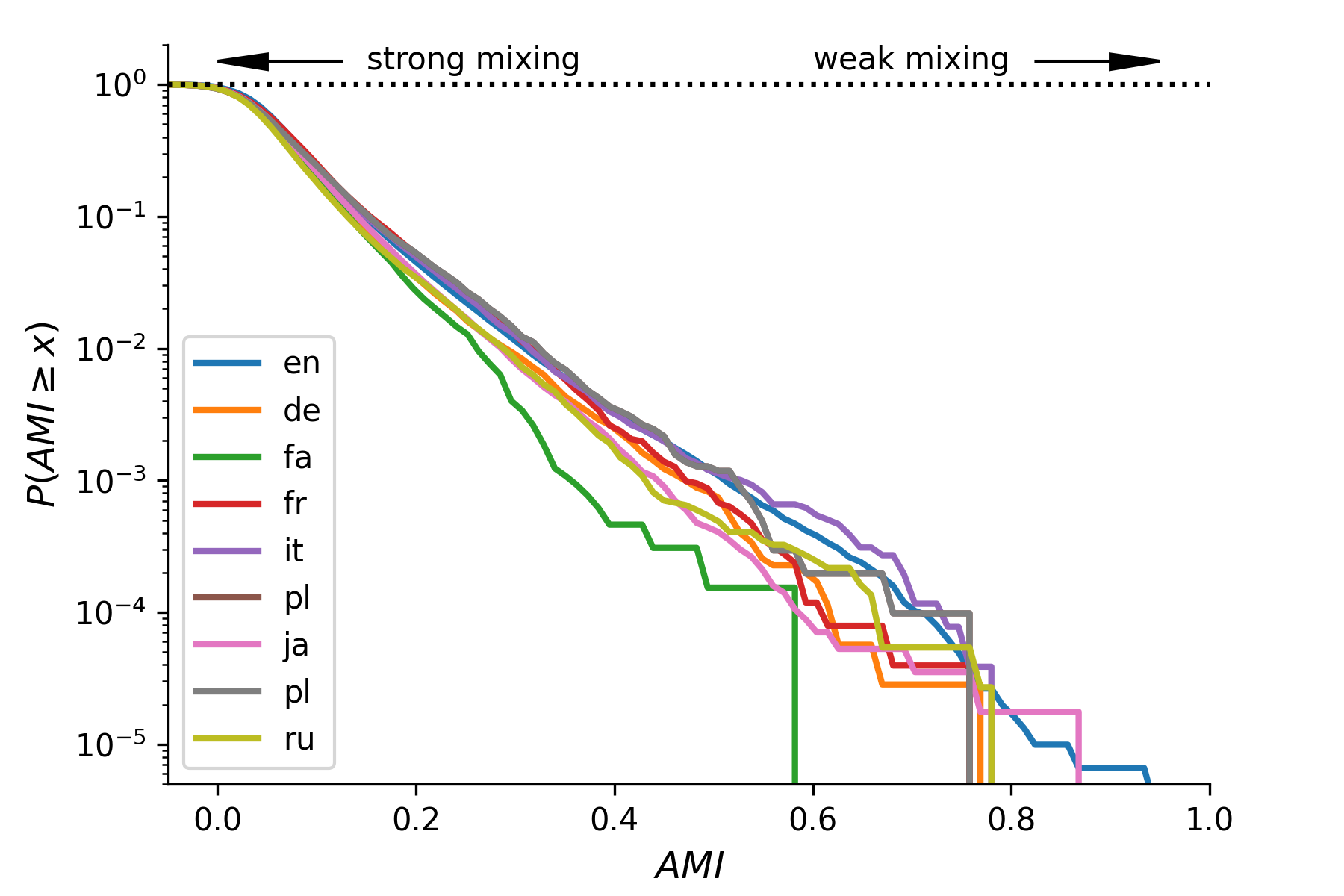} 
    \includegraphics[width=0.34\linewidth]{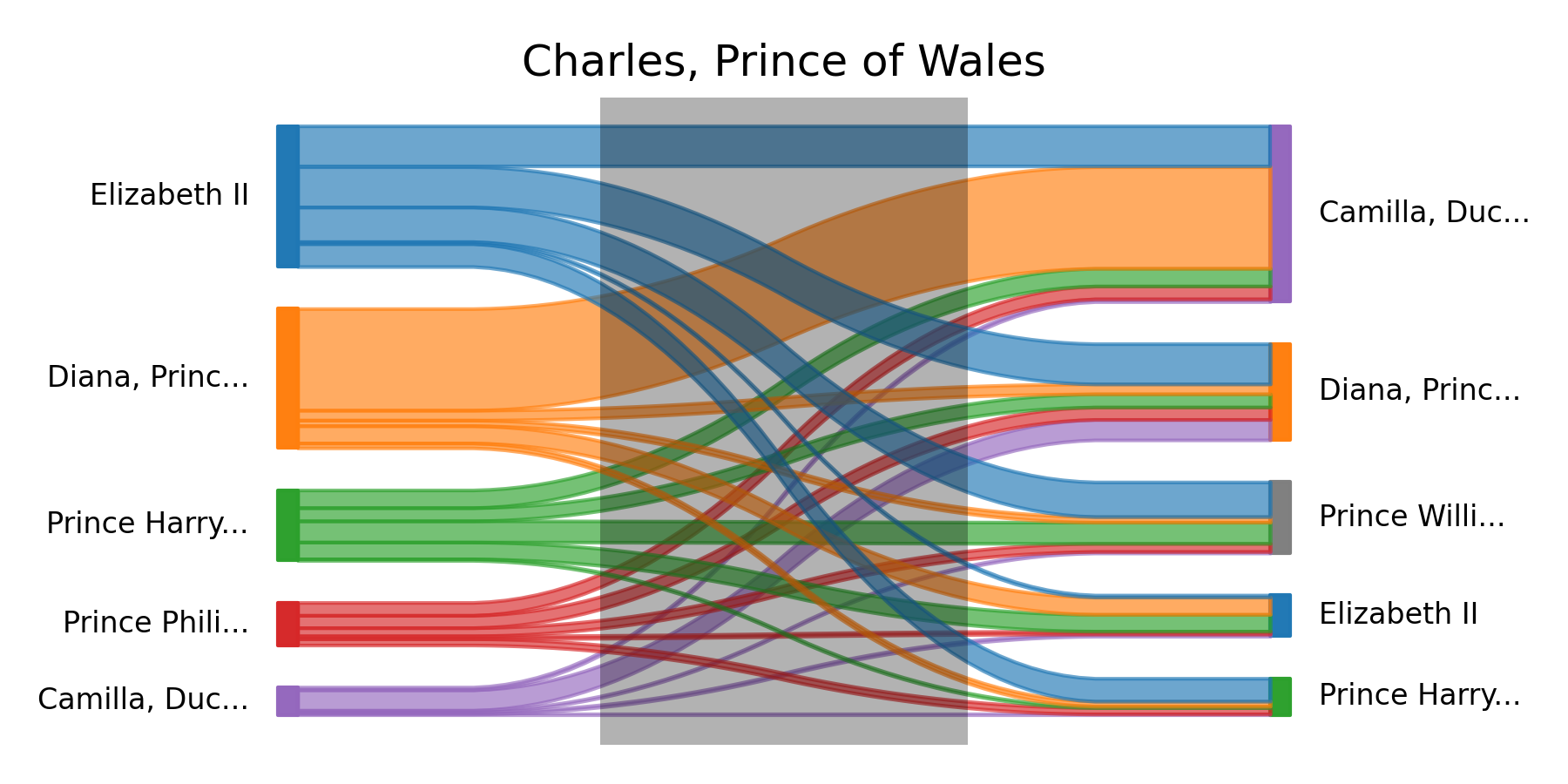} 
      \includegraphics[width=0.34\linewidth]{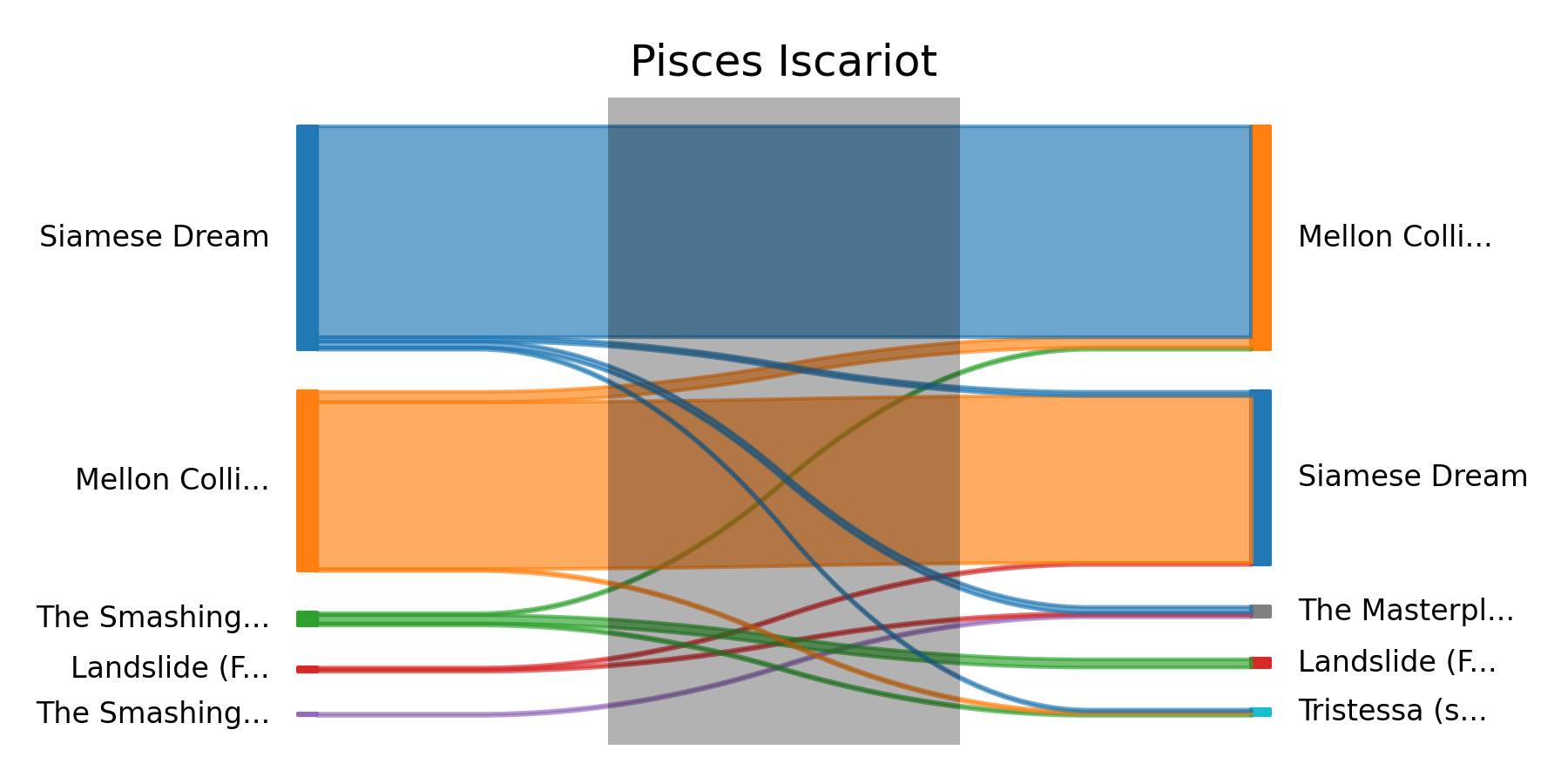} 
  \figcaption{Mixing of flows: Cumulative distribution of AMI values for all pages $m$ with more than $100$ sequences passing through them (left). Examples of sequences with strong mixing (middle, $AMI\approx0.1$) and weak mixing (right, $AMI\approx0.6$)}
  \vspace{1mm}
  \label{fig:mixing_ami_dist}
\end{figure*}

\xhdr{Semantic embedding} The semantic embeddings are learned from textual content of Wikipedia articles, thereby possessing the ability to capture the inter-article semantic similarity. The representation for each article in a given language is obtained by averaging the representations of the words---using pre-trained FastText~\cite{Grave2018learning} word embeddings with $300$ dimensions in the respective language~\cite{ft}---appearing in the first paragraph of the article. 
Similar to Sec.~\ref{sec:synthetic_gen}, we use the XML dump~\cite{wikipedia_xmldumps} dated `2021-04-01'. 
We use this resource for performing the semantic diffusion analysis in Sec.~\ref{sec:semantic_diffusion}.

\xhdr{Navigation embedding} As the name implies, the navigation embeddings are learned from the navigation sequences described in Secs.~\ref{sec:real_nav} and~\ref{sec:synthetic_gen}, thereby possessing the ability to capture semantic relatedness between the concepts that different articles represent. 
Analogous to text-based embeddings, the navigation sequences (sentences) are ordered collections of pageviews of articles (words). 
That said, following convention in the network representation learning literature \cite{deepwalk}, we train $128$-dimensional article embeddings from each of the real and synthetic navigation sequence datasets---with sequences of at least 2 pageviews---using FastText with the default hyperparameter settings prescribed in~\cite{Bojanowski2017enriching}. 
We refrain from hyperparameter tuning, since i) it captures the typical use-case, and ii) we are not interested in the absolute performance but only in the relative difference of the embeddings generated from the respective datasets. 
We use this resource for the semantic similarity/relatedness (Sec.~\ref{subsec:relatedness}) and topic classification (Sec.~\ref{subsec:topicpred}) tasks.

\xhdr{Added-links data} This data consists of new links added to Wikipedia in April 2021. 
Following the procedure described in~\citeauthor{ParanjapeImproving}, we obtain the added links by computing the set difference of links existent in Wikipedia in April and March 2021, respectively. Next, we restrict ourselves to added links possessing at least $10$ indirect paths between them observed in the real navigation sequences. 
We denote this set of links as positive examples $(s,t)\in \mathcal{L}$.
We also identify a set of negative examples $\tilde{\mathcal{L}}$, \ie, a set of links that were not added in the corresponding period. Please see Appendix~\ref{app:added_links_data} for details around the aforementioned construction. This dataset is used for the link prediction task in Sec.~\ref{subsec:linkpred}.

\xhdr{Similarity/Relatedness data}
We use the WikiSRS data \cite{wikisrs_data} as a ground-truth for capturing the relationships between articles. It contains the similarity and relatedness judgments (on a scale of $0 \ldots 100$) for 688 pairs of Wikipedia entities (people, places, and organizations), as assigned by 5 different Amazon Mechanical Turk workers. 
We map the entities to the corresponding language version of Wikipedia using the sitelinks from its Wikidata-item. As a result, a pair is not available for evaluation in a given language if at least one of the articles does not exist in that language.
We use this dataset for evaluating the navigation sequence-based article representations through the semantic relatedness task in Sec.~\ref{subsec:relatedness}.

\xhdr{Topic-labels data}
As a ground-truth for topic assignments to Wikipedia articles, we use the dataset described by Johnson \etal~\cite{LanguageAgnosticORES}.
A Wikipedia article can belong to one or more of 64 topics (such as `Mathematics', `Entertainment', `Politics and Government', \etc). The annotations come from editors interested in specific topics who are organized in the so-called WikiProjects and manually label relevant articles. The more than $2000$ different WikiProjects labels are aggregated to a set of 64 topics using a taxonomy~\cite{wikitax} derived from a hierarchy developed by the editor community~\cite{wp_council}. We extract WikiProjects labels for articles in English Wikipedia (almost all articles have at least one label). For articles in other languages, we apply the labels from the corresponding article in English. This dataset is used in the topic classification task described in Sec.~\ref{subsec:topicpred}.

%% file: analysis.tex
\section{Empirical Characterization}
\label{sec:approach}
In this section, we empirically characterize two key aspects of real navigation sequences that help us to distinguish them from synthetic navigation sequences.
By construction, the synthetic sequences follow a Markov process of order 1, \ie when generating a synthetic sequence, the probability for the next article only depends on the current article and not on any of the previously visited articles. 
Therefore, our aim is to quantify the degree to which previously visited articles affect the probability for the next article in real navigation sequences, and thus, differs from model selection approaches that fit Markov models of fixed order~\cite{UsersReallyMarkovian, Singer2014detecting, Petrovic2020learning}.
First, we quantify the mixing of incoming and outgoing navigation flows when passing through a given article using information-theoretic measures (Sec.~\ref{sec:mixing}).
Second, we quantify how much faster synthetic navigation sequences diffuse in an abstracted semantic space using an article-embedding (Sec.~\ref{sec:semantic_diffusion}).

\subsection{Mixing of flows}
\label{sec:mixing}
We assess the degree to which the flow of real navigation sequences is  mixing when passing through a specific article~\cite{Rosvall2014memory}.
For navigation sequences passing through a given article $m$, we quantify the relation between incoming traffic (source $s$) and outgoing traffic (target $t$). By construction, the synthetic navigation sequences exhibit full mixing due to the fact that incoming and outgoing traffic are completely uncorrelated. 

We calculate the adjusted mutual information (AMI)~\cite{Vinh2009information}. For navigation sequences passing through $m$, the mutual information (MI) measures the average amount of information (in bits) about the target-page $t$ when knowing the source-page $s$. The advantage of AMI is that it is normalized between 0 (full mixing, no information about $t$) and 1 (no mixing, full information about $t$) and takes into account non-zero values of the MI from finite-size effects. Please see Appendix~\ref{app:ami} for the mathematical formulation of AMI.

\begin{figure}
  \centering
    \includegraphics[width=0.99\linewidth]{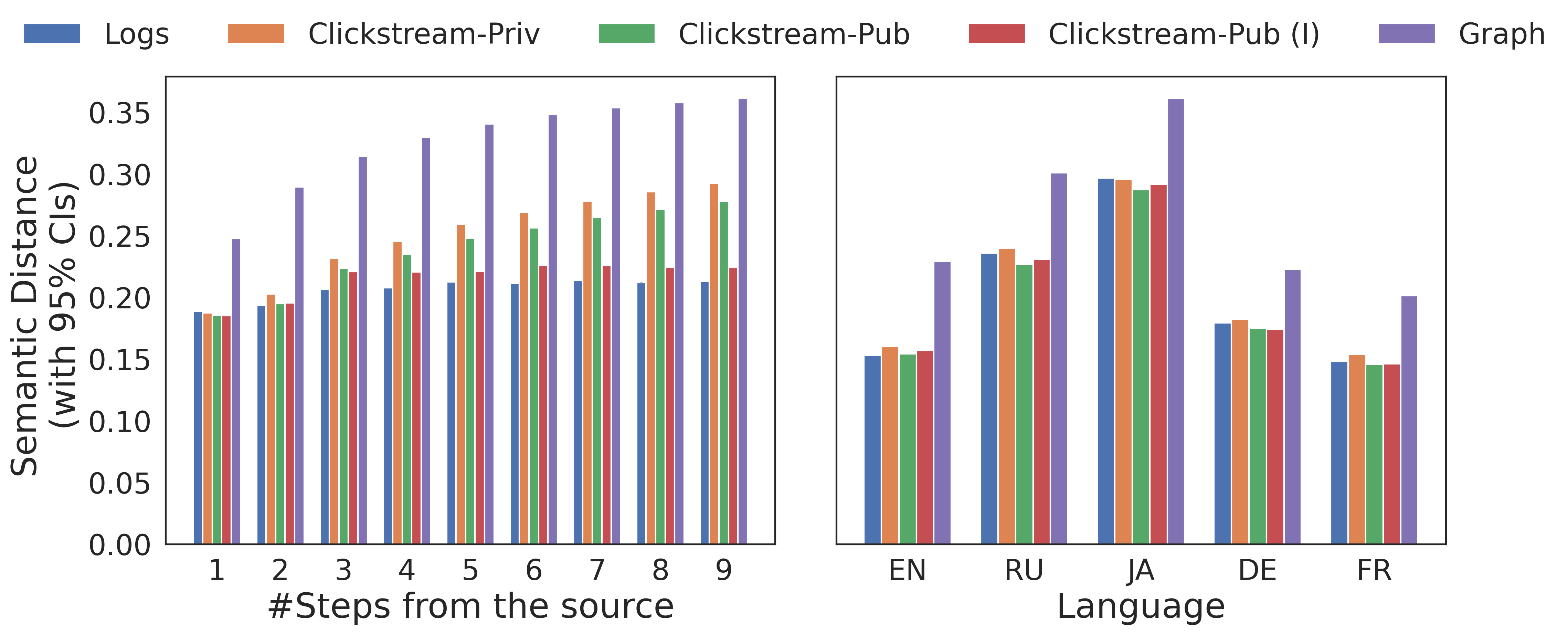}
    \figcaption{Diffusion in semantic space: Mean cosine distance between the first article and the article after (Left) $k$-steps ($1 \leq k \leq 9$) for \enshort; and (Right) $2$-steps for other languages.}
    \label{fig:semantic_diffusion}
\end{figure}

\martin{Fig.~\ref{fig:mixing_ami_dist} demonstrates that most pages show a very high degree of mixing: the distribution of AMI values is peaked around $0$ with an exponential decay and no further mode. 
In fact, less than $10\%$ of pages have an AMI $>0.1$ and fewer than $0.1\%$ of pages exhibit AMI $>0.6$.
In order to interpret these numbers, \Figref{fig:mixing_ami_dist} also shows two visual examples of pages with low and high AMI, respectively.
Qualitatively, we can see that for AMI $\approx0.1$ the flows overlap substantially such that starting from a given source $s$ (left), one can end up in virtually any of the targets $t$ (right) with similar probability when passing through $m$ (middle). In contrast, for AMI $\approx0.6$ flows virtually do not overlap such that choosing the source $s$ strongly determines the target $t$.}
We find a similar pattern for all 8 languages.
We find no statistically significant correlation between the number of observations (i.e. number of sequences passing through an article) and the value of AMI (Spearman's $\rho=0.0005$, $p=0.80$).

\subsection{Diffusion in semantic space}
\label{sec:semantic_diffusion}
Next, we assess whether a similar effect holds for sequences more generally, not just focusing on groups of sequences passing through a specific article.
For this, we quantify the diffusion of navigation sequences in a low-dimensional vector space obtained from the text of the articles. 
For the latter, we use the \textit{Semantic embedding} described in detail in Sec.~\ref{sec:data_auxiliary} capturing how semantically similar articles are.
We then map the navigation sequence into this space and measure the cosine-distance between the starting article and the article reached after $k$ steps in Fig.~\ref{fig:semantic_diffusion}. 
\martin{Increasing the step size $k$, we see that the average distance increases for the synthetic clickstream sequences, however, staying close to the value observed from the real navigation sequences. As stated in Sec.~\ref{sec:intro}, while all the differences are statistically significant (measured using bootstrapped 95\% confidence intervals), the effect sizes are small. Specifically, the difference between \logs and \cspub is $<10\%$ until $k=4$, and only for $k>5$, we observe larger effect sizes, however, we note that those sequences are very rare. Additionally, these findings hold across different languages. Moving beyond averages, we also analyze the entire distribution of semantic distances in Fig.~\ref{fig:semantic_diffusion_app}. Please see Appendix~\ref{app:diffusion} for details.}

%% file: downstream_tasks.tex
\begin{table*}[t]
\vspace*{1.5mm}
\tabcaption{\martin{Quantitative analysis for assessing the utility of five different types of navigation sequences across the eight Wikipedia language versions considered in this study using downstream tasks: (a) Next-article prediction, (b) Semantic relatedness/similarity, and (c) Topic classification. The best performance (higher numbers are better) is shown in \textbf{bold}.}}   
\label{tab:downstream_tasks}
\centering

\begin{threeparttable}

\subfloat[Next-article prediction (MRR)]{
  \centering
  \resizebox{0.9\linewidth}{!}{
    \begin{tabular}{llcccccccc|cccccccc}
    \toprule
    \multirow{2}{*}{} & \multirow{2}{*}{} & \multicolumn{8}{c|}{\bf All Queries} & \multicolumn{8}{c}{\bf Filtered Queries} \\
    \cmidrule(lr){3-10}
    \cmidrule(lr){11-18}
    \bf Type & \bf Dataset & \bf \enshort & \bf \jashort & \bf \deshort & \bf \rushort & \bf \frshort & \bf \itlshort & \bf \plshort & \bf \fashort & \bf \enshort & \bf \jashort & \bf \deshort & \bf \rushort & \bf \frshort & \bf \itlshort & \bf \plshort & \bf \fashort \\
    \midrule
    Real & \logs & \bf 0.369 \tnote{\dag} & \bf 0.315 \tnote{\dag} & \bf 0.275 \tnote{\dag} & \bf 0.317 \tnote{\dag} & \bf 0.316 \tnote{\dag} & \bf 0.347 \tnote{\dag} & \bf 0.302 \tnote{\dag} & \bf 0.388 \tnote{\dag} & \bf 0.595 \tnote{\dag} & \bf 0.615 \tnote{\dag} & \bf 0.646 \tnote{\dag} & \bf 0.644 \tnote{\dag} & \bf 0.690 \tnote{\dag} & \bf 0.686 \tnote{\dag} & \bf 0.693 \tnote{\dag} & \bf 0.666 \tnote{\dag} \\
    \hdashline
    Synthetic & \cspriv & 0.325 & 0.273 & 0.249 & 0.286 & 0.279 & 0.307 & 0.277 & 0.361 & 0.541 & 0.557 & 0.593 & 0.587 & 0.625 & 0.618 & 0.634 & 0.623 \\
    Synthetic & \cspub & 0.316 & 0.258 & 0.238 & 0.259 & 0.266 & 0.278 & 0.247 & 0.270 & 0.541 & 0.561 & 0.592 & 0.589 & 0.629 & 0.618 & 0.641 & 0.642 \\
    Synthetic & \cspub (I) & 0.288 & 0.222 & 0.197 & 0.214 & 0.212 & 0.236 & 0.191 & 0.221 & 0.537 & 0.557 & 0.591 & 0.586 & 0.625 & 0.618 & 0.642 & 0.639 \\
    Synthetic & \graph & 0.017 & 0.017 & 0.019 & 0.024 & 0.015 & 0.020 & 0.029 & 0.050 & 0.018 & 0.022 & 0.026 & 0.028 & 0.020 & 0.024 & 0.040 & 0.062 \\
    \bottomrule
    \end{tabular}
    \label{tab:next_article_pred}
  }
}
\\
\moveup
\subfloat[Semantic relatedness/similarity (Spearman's $\rho$)]{
  \resizebox{0.9\linewidth}{!}{
    \begin{tabular}{llcccccccc|cccccccc}
    \toprule
    \multirow{2}{*}{} & \multirow{2}{*}{} & \multicolumn{8}{c|}{\bf Relatedness} & \multicolumn{8}{c}{\bf Similarity} \\
    \cmidrule(lr){3-10}
    \cmidrule(lr){11-18}
    \bf Type & \bf Dataset & \bf \enshort & \bf \jashort & \bf \deshort & \bf \rushort & \bf \frshort & \bf \itlshort & \bf \plshort & \bf \fashort & \bf \enshort & \bf \jashort & \bf \deshort & \bf \rushort & \bf \frshort & \bf \itlshort & \bf \plshort & \bf \fashort \\
    \midrule
    Real & \logs & 0.769 & 0.728 & 0.693 & \bf 0.704 & 0.697 & \bf 0.710 & 0.691 & 0.665 & 0.722 & \bf 0.703 & 0.648 & \bf 0.677 & 0.662 & 0.665 & 0.630 & 0.621 \\
    \hdashline
    Synthetic & \cspriv & 0.764 & 0.689 & 0.673 & 0.688 & 0.714 & 0.703 & \bf 0.700 & 0.595 & 0.711 & 0.662 & 0.633 & 0.623 & \bf 0.672 & 0.652 & \bf 0.637 & 0.539 \\
    Synthetic & \cspub & 0.749 & 0.625 & 0.653 & 0.655 & 0.647 & 0.643 & 0.597 & 0.541 & 0.703 & 0.626 & 0.621 & 0.612 & 0.620 & 0.619 & 0.520 & 0.530 \\
    Synthetic & \cspub (I) & 0.715 & 0.619 & 0.632 & 0.613 & 0.586 & 0.592 & 0.592 & 0.480 & 0.653 & 0.571 & 0.574 & 0.573 & 0.513 & 0.540 & 0.530 & 0.444 \\
    Synthetic & \graph & \bf 0.771 & \bf 0.750 & \bf 0.709 & 0.685 & \bf 0.723 & 0.703 & 0.691 & \bf 0.677 & \bf 0.734 & 0.691 & \bf 0.674 & 0.638 & 0.661 & \bf 0.666 & 0.619 & \bf 0.633 \\
    \bottomrule
    \end{tabular}
  }
  \label{tab:relatedness}
}
\\
\moveup
\subfloat[Topic classification (\fscore)]{
  \resizebox{0.9\linewidth}{!}{
    \begin{tabular}{llcccccccc|cccccccc}
    \toprule
    \multirow{2}{*}{} & \multirow{2}{*}{} & \multicolumn{8}{c|}{\bf Micro Aggregates} & \multicolumn{8}{c}{\bf Macro Aggregates} \\
    \cmidrule(lr){3-10}
    \cmidrule(lr){11-18}
    \bf Type & \bf Dataset & \bf \enshort & \bf \jashort & \bf \deshort & \bf \rushort & \bf \frshort & \bf \itlshort & \bf \plshort & \bf \fashort & \bf \enshort & \bf \jashort & \bf \deshort & \bf \rushort & \bf \frshort & \bf \itlshort & \bf \plshort & \bf \fashort \\
    \midrule
    Real & \logs & \bf 0.628 \tnote{\dag} & \bf 0.667 & 0.621 & \bf 0.633 \tnote{\dag} & \bf 0.618 & \bf 0.623 & 0.633 & 0.589 & \bf 0.569 \tnote{\dag} & \bf 0.567 \tnote{\dag} & \bf 0.547 \tnote{\dag} & \bf 0.563 \tnote{\dag} & \bf 0.560 \tnote{\dag} & \bf 0.557 \tnote{\dag} & \bf 0.541 \tnote{\dag} & 0.496 \\
    \hdashline
    Synthetic & \cspriv & 0.597 & 0.646 & 0.595 & 0.609 & 0.589 & 0.594 & 0.609 & 0.571 & 0.544 & 0.547 & 0.523 & 0.539 & 0.532 & 0.531 & 0.512 & 0.477 \\
    Synthetic & \cspub & 0.586 & 0.618 & 0.575 & 0.586 & 0.556 & 0.562 & 0.587 & 0.549 & 0.531 & 0.513 & 0.491 & 0.506 & 0.496 & 0.489 & 0.478 & 0.446 \\
    Synthetic & \cspub (I) & 0.524 & 0.561 & 0.495 & 0.522 & 0.449 & 0.461 & 0.502 & 0.453 & 0.464 & 0.431 & 0.396 & 0.436 & 0.378 & 0.375 & 0.387 & 0.335 \\
    Synthetic & \graph & 0.625 & 0.666 & \bf 0.636 \tnote{\dag} & 0.628 & \bf 0.625 \tnote{\dag} & 0.621 & \bf 0.639 & \bf 0.600 \tnote{\dag} & 0.563 & 0.543 & 0.535 & 0.547 & 0.555 & 0.543 & 0.526 & \bf 0.499 \\
    \bottomrule
    \end{tabular}
  }
  \label{tab:topic_pred}
}
\moveup
\begin{tablenotes}
\item[\dag] Indicates statistical significance ($p<0.05$) between the best and the second-best method using bootstrapped 95\% confidence intervals.
\end{tablenotes}
\end{threeparttable}%
\moveup
\moveups
\end{table*}

\section{Downstream Tasks}
\label{sec:tasks}
The empirical observations in the previous section suggest that synthetic navigation sequences from clickstream data are very similar to real navigation sequences. 
In this section, we investigate the implications of the aforementioned finding for practical downstream applications. 
For this, we consider four tasks relevant in the context of reader navigation or for which navigation sequences have been shown to provide useful signals: 
predicting the next article in the navigation sequence (Sec.~\ref{subsec:nextarticle}), predicting new links to be added (Sec.~\ref{subsec:linkpred}), inferring semantic similarity/relatedness between entities (Sec.~\ref{subsec:relatedness}), and classifying articles into topics (Sec.~\ref{subsec:topicpred}).
In each case, the goal is not to propose new methods to achieve state-of-the-art performance in these tasks.
Rather, we are interested in assessing the relative difference in performance when using synthetic instead of real navigation sequences.

\subsection{Next-article prediction}
\label{subsec:nextarticle}
We predict the next article in the navigation sequence following typical setup in sequence-based recommendation tasks~\cite{Quadrana2018recommender}. 
Training a model on each of the different datasets (real and synthetic navigation sequences), we evaluate their ability to predict the next article in real navigation sequences.
The main idea is that differences between training and test data should be reflected in a decrease in the prediction score. 

Specifically, we use triples of consecutively read articles, $(s_1, s_2, t)$, sampled uniformly from the navigation sequences.
For the training, we select $80\%$ of the triples as training data and train a separate Markov chain model of order 2 for each of the 4 datasets, respectively. 
For the evaluation, we select $10\%$ of the triples each for the validation- and test-set from the real navigation sequences for all four cases.
For each triple $(s_1,s_2,t=t^*)$ in the test-set, we use the trained model to compute a ranked list of candidate targets linked from the sources $(s_1, s_2)$ using their predicted likelihoods $P(t|s_1,s_2)$, and then determine the rank of the true target $r(t^*)$. We evaluate the predictions using mean reciprocal rank (MRR) (we also measured Recall@k yielding similar results, not shown).

In Table~\ref{tab:next_article_pred} we observe that, as expected, training the model on synthetic sequences leads to a decrease in the prediction score. 
While the performance using synthetic sequences from the unbiased random walk is extremely poor, synthetic sequences from clickstream yield an MRR within a relative difference of $10-20\%$ when compared to the real navigation sequences. 
Interestingly, for languages with less data (in terms of the number of navigation sequences), the performance of the public clickstream becomes substantially worse than the private clickstream. 
This suggests that $k$-anonymity (only links with more than $10$ clicks are included in \cspub) plays a larger role than the restriction to first-order transitions.
This is further corroborated when looking at a set of filtered queries, where we prune all queries that lack observations in the training set for any one of the four types of navigation sequences. 
In addition to yielding overall higher (almost double) values of MRR, the difference between private and public clickstream virtually disappears.

\subsection{Link prediction}
\label{subsec:linkpred}
We predict new links $(s,t)$ in the Wikipedia graph from a source page $s$ to a target page $t$ following the approach by \citeauthor{ParanjapeImproving}~\cite{ParanjapeImproving}, who showed that navigation sequences contain a useful signal for the prediction of new links in Wikipedia. 
From the \textit{Added links data} we obtain a labeled dataset of links consisting of positive $\mathcal{L}$ examples (links that were added) and negative $\bar{\mathcal{L}}$ examples (links that were not added).
Using real and synthetic navigation sequences, respectively, we  calculate the path proportion for each link $(s,t)\in \mathcal{L},\bar{\mathcal{L}}$ according to
$p(s,t) = N(s,t)/N(s)$, 
where $N(s,t)$ is the number of sequences from $s$ to $t$ and $N(s)$ are all sequences starting in $s$.   
We rank all links according to $p(s,t)$ in descending order and calculate the precision@k, i.e. what fraction of the top-k ranked links correspond to positive examples.

\begin{figure*}
	\moveup
    \includegraphics[width=0.75\linewidth]{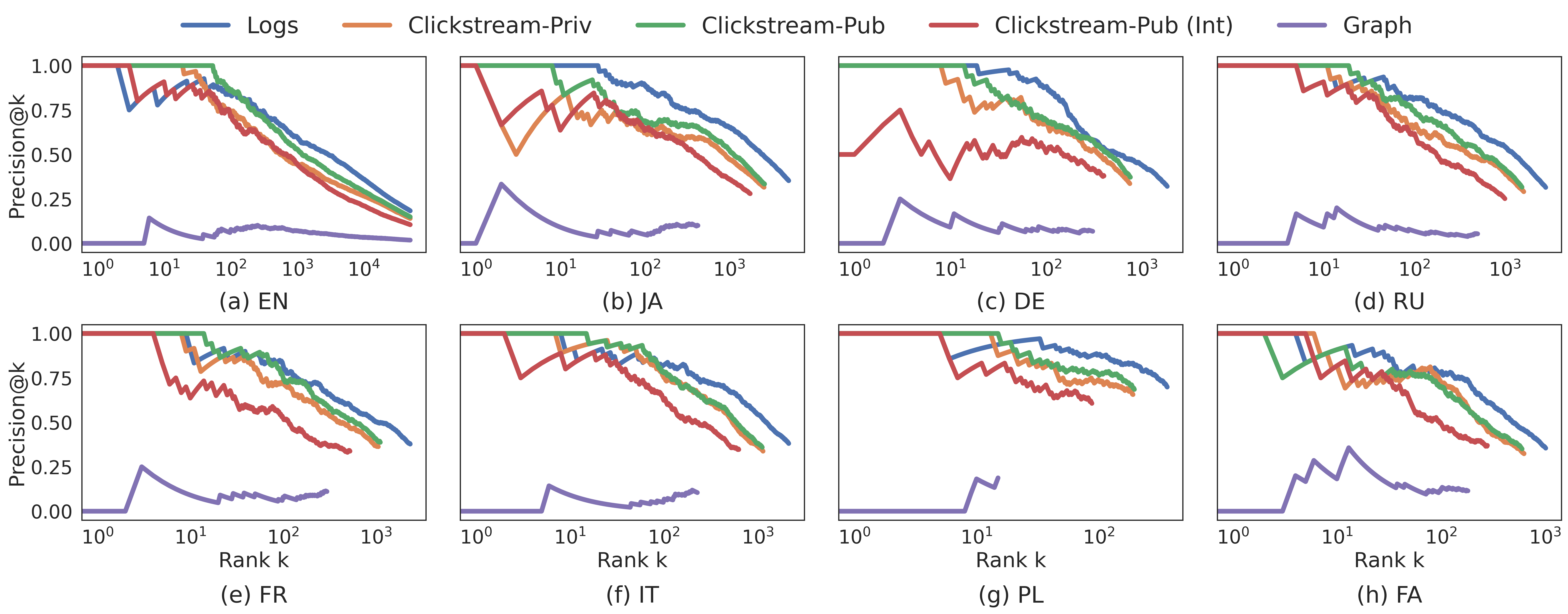}
    \moveups
    \figcaption{Link prediction task: Performance measured using precision@k
    on five different types of navigation sequences across the eight Wikipedia language versions considered in this study. Predictions cannot be made (indicated by truncated precision curves) for links where no path was observed in the navigation sequences.}
    \label{fig:link_pred}
\end{figure*}

In Fig.~\ref{fig:link_pred} we see that precision@k decreases with $k$ in most cases from $1$ ($k=1$) to $\approx0.4$ ($k=1000$). The only exception are the predictions from the synthetic sequences based on the unbiased random walk, which yields a poor performance.
More importantly, the performance from the synthetic sequences based on the clickstream is comparable with the real navigation sequences for all values of $k$. In most cases the relative difference is less than $10\%$.
\martin{Interestingly, for small $k$ \cspub performs slightly better than \cspriv, possibly due to filtering of negative low-probability pairs from the k-anonymity threshold for the former. }

\subsection{Semantic similarity/relatedness}
\label{subsec:relatedness}
We evaluate how well navigation sequences can be used to infer relationship between entities.
Following the approaches in~\cite{SemanticRelatednessHumanNavigational, Dallmann2016extracting}), we first generate an embedding of articles from navigation sequences for the real and the synthetic datasets, respectively.
We then compare the cosine-similarity between articles in this representation with a ground-truth dataset (\textit{Similarity/Relatedness data}) containing human-annotated ratings for similarity and relatedness for a selected set of pairs.
Specifically, we calculate Spearman's $\rho$ between the two ranked lists of pairs of articles.

In Table~\ref{tab:relatedness} we see that the synthetic sequences from \cspriv yield values for $\rho$
that are compatible with the ones obtained from real navigation sequences.
In contrast, values for $\rho$ from the synthetic sequences from \cspub are substantially lower, though the relative difference is still well within $10-20\%$.
Similar to the task of next-article prediction, this suggests that the $k$-anonymity (only links with more than $10$ clicks are included in \cspub) play a larger role than the restriction to first-order transitions. 
This is further corroborated by the results of the synthetic navigation sequences from the unbiased random walk which yield values for $\rho$ that are similar or sometimes better than the real navigation sequences. 
We note that this finding is consistent with previous findings reported in~\cite{SemanticRelatednessHumanNavigational, Dallmann2016extracting}) and that the reported values of $\rho$ are on-par with the state-of-the-art results from the study introducing the ground-truth dataset~\cite{semantic_relatedness}. 
One interpretation of these observations is that many of the existing links, even if they are rarely or never used, are crucial for capturing the complexity in the relationships between entities when learning representations. 

\subsection{Topic classification}
\label{subsec:topicpred}
We predict topics of articles in a supervised classification task using the embeddings obtained from the different navigation sequences. We follow the approach described in~\cite{Asthana2018feweyes,LanguageAgnosticORES}, who used embeddings generated from the content of the articles (text or links) as features to predict labels from topic annotations of editors.

We use (\textit{Navigation embeddings}) as features, i.e. the embeddings generated from the real and synthetic navigation sequences, respectively.
We perform supervised classification of the topic labels from the \textit{Topic-labels data}, where an article can have one or more of the $64$ different topic labels.
For each topic label we train an independent binary classifier using logistic regression, for which we use the publicly available implementation in \sklearn~\cite{sklearn} with the prescribed default parameters. 
We use $80\%$ as the training set, $10\%$ as validation, and $10\%$ as the test set. 
We report micro and macro-F1 statistics.
We explored other statistics (precision and recall) as well as the effect of hyperparameters  on the resultant performance, specifically, the regularization parameter $\alpha$ and the usage of reciprocal class frequencies as weights in the loss computation for managing the class imbalance. 
The results portrayed similar trends, and are therefore omitted.

In Table~\ref{tab:topic_pred} we observe that the performance for topic classification does not vary strongly between real and synthetic datasets as features.
Using navigation embeddings generated from public clickstream data yields the lowest scores.
However, the relative difference to using real navigation sequences remains well below $10\%$ in most cases. 
Surprisingly, embeddings generated from the unweighted hyperlink graph yield a performance on par with real navigation sequences.
Overall, these results largely mirror the findings from the downstream task on semantic similarity/relatedness in the previous section (Sec.~\ref{subsec:relatedness}).
The latter also used embeddings generated from navigation sequences.
Therefore, we can conclude that article embeddings generated from synthetic navigation sequences are of comparable quality to those generated from real navigation sequences.

%% file: discussion.tex
\section{Discussion and Concluding Insights}
\label{sec:discussion}

\subsection{Summary of results}
\xhdr{Characterizing the flow of navigation sequences}
Across all 8 considered languages, our results show that navigation sequences passing through a given article are strongly mixing with most pages exhibiting a mutual information close to zero, i.e. knowing the source-article provides little information about the target-article of the navigation sequence.
However, for a small set of articles (around $0.1-1\%$) we observed much larger values of mutual information, which provides clear evidence for cases where the real navigation sequences differ substantially from the synthetic navigation sequences.
We showed that, as a result, overall the diffusion in the semantic space of articles is virtually indistinguishable when comparing real navigation sequences with synthetic sequences from public clickstream.
While differences in the cosine-similarity between articles separated by $k$ steps are statistically significant, the effect sizes are small (Table~\ref{app:relative_difference}). 

\xhdr{Performance of synthetic navigation sequences in downstream tasks}
We showed that synthetic navigation sequences from the public clickstream dataset are effective for many practical applications for all 8 considered languages.
We compared the performance of real and synthetic navigation sequences in four different downstream tasks involving the use of navigation sequences ( link prediction~\cite{ParanjapeImproving}, next-article prediction~\cite{Cordonnier2019extrapolating}, generating representations~\cite{Sen2019towards}, and topic classification~\cite{LanguageAgnosticORES}), 
revealing that using clickstream data often yields performance that are within $10\%$ (or less) in comparison to using real navigation sequences (Table~\ref{app:relative_difference}). 
Specifically, article embeddings generated from synthetic navigation sequences are of comparable quality to those generated from real navigation sequences.
Furthermore, in the case of next-article prediction and semantic similarity/relatedness we found evidence that the main limitations do not originate from the restrictions to first-order Markov processes when generating the synthetic data but from additional filtering of the public data using $k$-anonymity necessary for ensuring privacy.

\subsection{Implications and Future Work}
\xhdr{When synthetic data is enough}
Our results indicate that, for many practical cases, the synthetic navigation sequences from public clickstream data can be used as a good approximation for real navigation sequences in Wikipedia. 
\martin{In order to generate the synthetic sequences in practice, one cannot exactly match length and starting articles of each real sequence; it is, however, possible to match these properties on average by sampling from the distributions of lengths~\cite{ImprovingHumanNavigation} and overall pageview statistics~\cite{wikipedia_pageview_dumps}, respectively. Alternatively, it is possible to circumvent the requirement of matching on length by using intrinsic stopping similar to \cspub (I) style sequences.}
For cases in which necessary privacy filters (\eg, $k$-anonymity) are believed to remove too many rarely-used links, one potential solution would be to augment the public clickstream data with the unweighted hyperlink graph.

There are several implications of this finding.
First, previous research investigating navigation on Wikipedia using the public clickstream data (e.g. on search strategies~\cite{SearchStrategies}) could be generalized to describe navigation on Wikipedia more generally; naturally, it is unlikely to capture all the nuances of how readers browse Wikipedia \cite{piccardi2021largescale}. 
Second, research on navigation in Wikipedia becomes more accessible to a wider range of researchers as the data underlying the synthetic sequences is publicly available for 11 languages and updated on a monthly basis. 
Other resources such as article representations learned from navigation embeddings, such as the so-called navigation vectors~\cite{wiki_nav_vectors}, can now be generated from publicly available datasets and thus become much more widely available and customizable. 
Third, our results question the un-scrutinized approaches to improving our understanding of user behavior through more and more data which often comes at significant costs to user privacy. 
\martin{The example of navigation in Wikipedia provides clear evidence that for many practical downstream tasks (of which we gave four examples) the available synthetic data can be considered ``good enough''}. 
Fourth, the Wikipedia clickstream data provide an example case for how to approach research on navigation of users in online platforms more generally in a privacy-preserving way. 
It remains an open question whether our findings will generalize beyond Wikipedia.
However, our results highlight that publicly sharing clickstream-like data constitutes an immensely valuable data source and empowers others to better understand navigation on these platforms. 

\xhdr{When synthetic data is not enough}
\martin{The clickstream represents an excellent resource to approximate the global behavior of the readers on Wikipedia, but it cannot replace the server logs entirely. The fine-grained private activities stored on the servers, despite the careful anonymization guidelines of the Wikimedia Foundation, remain a valuable resource for studies that focus on the properties of readers. Real navigation sequences could help answer questions that rely on keeping track of the activities of the same user, such as revisitation patterns, multi-tab behavior, and how readers interact with additional content on a Wikipedia page (e.g., images or infoboxes \cite{wiki_infobox}). At the same time, features removed during the aggregation of the public clickstream, such as time and geolocation, can support a more in-depth understanding of the information consumption patterns of Wikipedia readers \cite{piccardi2021largescale}. To this end, an interesting future direction would be to build generative models capable of generating realistic navigation sequences in a privacy-preserving manner.
}

\xhdr{Memory in navigation sequences}
\martin{
The comparison between real and synthetic data provides insights on the degree of memory in navigation sequences; \ie, we characterize memory not at the level of an individual sequence but as a collective property emerging from averaging over hundreds of millions of sequences. By construction, the synthetic sequences are generated from a process without memory where the next step does not depend on any of the previously visited pages, such that deviations from the real sequences indicate importance of memory.
This problem is typically approached by estimating the order of a Markov-process fitted to the data for, e.g., Web~\cite{UsersReallyMarkovian} or targeted navigation~\cite{Singer2014detecting}. While previous works have yielded inconclusive results, our results suggest that fitting a single order constitutes an ill-posed problem.
While we find that most paths do not possess memory (\cf Fig.~\ref{fig:mixing_ami_dist} where most pages are characterized by mutual information close to $0$), we cannot claim that readers follow a Markovian process of order $1$. In fact, we note that there could be a plethora of small but consequential subsets of navigation sequences with extremely strong memory (\eg, a reporter researching a topic for their front-page article). 
}

\xhdr{Context of readers during navigation}
\martin{Our findings serve as a starting point for more in-depth studies taking into account the context of the reader by providing ground-truth labels collected from, e.g., surveys. In fact, previous research on why readers visit Wikipedia~\cite{singer_why_2017, lemmerich_why_2019} showed that motivations are diverse (\eg `intrinsic learning', `bored', \etc), or that some topics are more viewed in some regions (\eg STEM in countries with lower human development index). This suggests that future work should aim at better understanding the heterogeneity of readers and context of their visit. Furthermore, one promising direction would be to study reading behavior on longer timescales identifying learning pathways in order to, e.g., generate curricula around broader topics~\cite{Sayyadiharikandeh2019finding}.}

\subsection{Methodological limitations}

\xhdr{Pageload as a proxy for reading}
The navigation sequences only capture pageloads from the http-request to Wikimedia's server. 
Our data does not capture actual reading behavior, such as how much time a reader was spending on a specific page~\cite{DwellingTime}.
Thus, the pageloads only serve as a proxy. 

\xhdr{Wikipedia readers}
The construction of the navigation sequences requires the grouping of pageloads by users.
For privacy reasons (e.g. no use of cookies), we use an approximate unique id by concatenating client IP address and user-agent  (\cf Sec.~\ref{sec:real_nav}). 
This can lead to aggregation of pageviews by different readers into the same sequence (e.g. through a shared network) or separation of pageview by the same reader into different sequences (e.g. by using several devices). 
Therefore, the attribution of pageviews to readers is imperfect. 
For more information on the implications of this approach, we refer to the detailed discussion by \citeauthor{singer_why_2017}
\cite{singer_why_2017} who use a similar methodology. 

%% file: appendix.tex
\appendix

\section{Implementation details}
\subsection{Constructing navigation sequences}
We mimic a reader via an MD5 digest of the concatenation of the client IP address and the user agent string, which serves as an approximate reader ID. The user agent information is also used to discard common bots and crawlers.

\subsection{Semantic embedding}
As stated in Sec.~\ref{sec:data_auxiliary}, semantic embeddings for each Wikipedia article is obtained by using the content in its abstract, \ie, all the content before the first section. To this end, we used the Python package mwtext (v0.0.1)~\cite{mwtext} for extracting the article abstracts from its wikitext-markdown.

\subsection{Added-links data}
\label{app:added_links_data}
Given all $s$ and $t$ appearing in a link $(s,t)\in\mathcal{L}$ we consider as negative examples all $(s,t')$ and $(s',t)$ for which i) $(s,t')$ or $(s',t)$ is not already a link, ii) $(s,t'),(s',t)\notin \mathcal{L}$, and iii) there are at least $10$ navigation sequences with an indirect path between $s$ and $t'$ or $s'$ and $t$, respectively. 

\begin{table}[!htb]
\centering
\caption{\label{tab:data_stats_app} Dataset statistics.}
\resizebox{0.99\linewidth}{!}{
    \begin{tabular}{lrrr}
    \toprule
    \textbf{Language} & \textbf{\#Pages} & \textbf{\#Links} & \textbf{\#NavigationSequences} \\\midrule    
    \en (\enshort) & 6,279,688 & 228,874,000 & 810,994,170 \\
    \ja (\jashort) & 1,261,262 & 59,872,582 & 122,157,015 \\ 
    \de (\deshort) & 2,557,540 & 76,790,852 & 89,441,061 \\
    \ru (\rushort) & 1,711,949 & 47,989,746 & 81,877,016 \\
    \fr (\frshort) & 2,314,313 & 73,343,009 & 70,321,709 \\
    \itl (\itlshort) & 1,683,856 & 48,708,959 & 65,974,945 \\
    \pl (\plshort) & 1,465,640 & 31,460,165 & 27,531,169 \\
    \fa (\fashort) & 778,667 & 9,660,952 & 15,845,197 \\
    \bottomrule
    \end{tabular}
    }
    \moveup
    \moveup
\end{table}

\subsection{Mixing of flows: AMI}
\label{app:ami}
Specifically, we get all triples of consecutively visited pages $(s,m,t)$.
For every distinct page $m^*$ we consider the set $\left\{   (s,m=m^*,t) \right\}$ of all triples passing through $m^*$ and calculate the joint distribution 
\begin{equation}
P_{m^*}(s=s^*,t=t^*)= \frac{ \left| \left\{   (s=s^*,m=m^*,t=t^*) \right\}\right| }{\left| \left\{   (s,m=m^*,t) \right\} \right|}
\end{equation} counting the fraction of triples starting in $s^*$ and ending in $t^*$ while passing through $m^*$. 
We calculate the AMI for each page $m$ to assess the degree of mixing of the trajectories passing through. 
We first calculate the MI between all sources $s\in S$ and targets $t\in T$ as  
\begin{equation}
MI_{m}(S,T)=\sum_{s\in S,t \in T} P_m(s,t) \log \frac{P_m(s,t)}{P_m(s)P_m(t)}
\end{equation}
from which we get the AMI as
\begin{equation}
AMI_m(S,T)=\frac{ MI_m(S,T)- \hat{MI}_m(S,T)}{\max{\left\{H(S),H(T)\right\}}-\hat{MI}_m(S,T)}.
\end{equation}
where $H(S), H(T)$ are the entropies over the marginal distributions and $\hat{MI}_m(S,T)$ is the expected MI from finite-size effects when randomly relating source and target. In practice, we use the implementation~\cite{ami_sklearn} from \sklearn~\cite{sklearn}. 

\section{Diffusion in semantic space}
\label{app:diffusion}
We observe in Fig.~\ref{fig:semantic_diffusion_app} that the distribution of distances for real and synthetic sequences from clickstream are almost completely overlapping (even for larger $k$) showing that differences are present but overall small. In contrast, the distribution of distances from the synthetic sequences of the unweighted random walk on the hyperlink graph is substantially shifted to larger values indicating that these sequences are much less confined and, as a result diffuse much faster. As a sanity-check, we show that the semantic distance between two completely randomly drawn articles is much larger corroborating that the embedding space captures the semantic similarity between articles.

\begin{table*}[t]
\caption{\martin{Quantifying the effect size measured as the percentage relative difference of the observed outcomes between real (\logs) and synthetic (\cspub) navigation sequences for the analyses and downstream tasks in this study. A negative relative difference indicates that synthetic navigation sequences (\cspub) achieve a better outcome than real navigation sequences (\logs).}}
\label{app:relative_difference}
\centering
\resizebox{0.9\linewidth}{!}{
    \begin{tabular}{lcccccccc}
    \toprule
    \multirow{2}{*}{} & \multicolumn{8}{c}{\bf Language version} \\
    \cmidrule(lr){2-9}
    \bf Task & \bf \enshort & \bf \jashort & \bf \deshort & \bf \rushort & \bf \frshort & \bf \itlshort & \bf \plshort & \bf \fashort \\
    \midrule
    Semantic distance ($k=1$) & -1.49 & -0.98 & -1.28 & -1.25 & -2.4 & -2.33 & -1.18 & -0.79 \\
    Semantic distance ($k=3$) & 11.1 & 2.32 & 6.43 & 6.21 & 8.17 & 12.96 & 4.09 & 5.44 \\
    Semantic distance ($k=5$) & 28.93 & 5.12 & 19.11 & 14.77 & 19.3 & 36.43 & 9.24 & 14.91 \\
    Semantic distance ($k=9$) & 68.59 & 10.11 & 47.99 & 23.72 & 29.4 & 58.57 & 24.49 & 43.6 \\
    \hdashline
    Next-article prediction (All queries) & 14.38 & 18.01 & 13.33 & 18.51 & 15.71 & 19.92 & 18.31 & 30.56 \\
    Next-article prediction (Filtered queries) & 9.20 & 8.85 & 8.32 & 8.60 & 8.86 & 9.93 & 7.58 & 3.64 \\
    \hdashline
    Semantic relatedness & 2.58 & 16.45 & 6.05 & 7.48 & 7.67 & 10.39 & 15.64 & 22.94 \\
    Semantic similarity & 2.61 & 12.19 & 4.38 & 10.64 & 6.86 & 7.47 & 21.30 & 17.18 \\
    \hdashline
    Topic classification (Micro) & 6.67 & 7.47 & 7.43 & 7.35 & 10.08 & 9.78 & 7.18 & 6.78 \\
    Topic classification (Macro) & 6.68 & 9.61 & 10.24 & 10.08 & 11.32 & 12.13 & 11.56 & 10.12 \\
    \hdashline
    Link prediction (P@10) & -25.00 & 10.00 & 0.00 & 0.00 & 0.00 & -11.11 & -11.11 & 0.00 \\
    Link prediction (P@50) & -13.64 & 17.78 & 18.75 & 13.04 & 0.00 & -4.55 & 11.11 & 2.50 \\
    Link prediction (P@100) & -2.38 & 22.47 & 20.45 & 7.41 & 8.43 & 4.88 & 12.50 & 10.26 \\
    \bottomrule
    \end{tabular}
}
\end{table*}